\documentclass[twocolumn,showpacs,showkeys,superscriptaddress]{revtex4}

\usepackage{amsmath}
\usepackage{bbold}
\usepackage{amsfonts}
\usepackage{amssymb}
\usepackage{pbsi}
\usepackage[T1]{fontenc}
\usepackage{hyperref}
\usepackage{xcolor}
\usepackage{graphicx}
\usepackage{subfig}
\usepackage{float}

\usepackage{color}

\begin{document}

\title{Different kinds of accelerated  propagation of relativistic electromagnetic plasma wavepackets}

\author{Felipe A. Asenjo}
\email{felipe.asenjo@uai.cl (corresponding author)}
\affiliation{Facultad de Ingenier\'ia y Ciencias,
Universidad Adolfo Ib\'a\~nez, Santiago 7491169, Chile.}

\begin{abstract}
Relativistic electromagnetic plasma waves are described by a dynamical equation that can be solved not only in terms of plane waves, but for several different accelerating wavepacket solutions. Depending on the spatial and temporal dependence of the plasma frequency, 
different kinds of accelerating solution can be obtained,  for example, in terms of Airy or Weber functions. Also, we show that an arbitrary accelerated wavepacket solution is possible, for example, for a system with a luminal plasma slab. 
\end{abstract}

\maketitle

\section{Introduction}

Accelerating wavepacket propagation has been shown to be a prolific field of study, spanning, for example,  from light \cite{neomi,Jiang,abdo,bouch,esat,panag,moya,Baumgartl,Nikolaos,chong, Kaminer} to fluids \cite{water}, sound \cite{chencho,zhao}, heat \cite{bookolivier,asenjoheat, asenjoheat2} or gravity \cite{asenjograv}. 
In plasma physics, accelerating wavepacket, in terms of Airy functions, has been found for  electron plasma waves \cite{li,asenjoelectron,YChen,LWu}, plasmons \cite{klimo} and for nonlinear waves in
weakly and strongly magnetized relativistic
plasmas \cite{asenjoelectron2}.

In this work we explore accelerating solutions for cold electron-positron relativistic electromagnetic plasma waves. They are described by a general (not perturbed) relativistic wave equation that can be obtained for a general plasma frequency, which can be dependent on space and time. In the case of constant plasma frequency, these modes are customary solved in terms of plane waves, with its corresponding dispersion relation associated to it. However,  we below show that also in such cases, accelerating Airy wavepacket modes also exist, where each part of the mode follow parabolic trayectories in the space formed by a transversal direction and one light-cone coordinate. Besides that, when other specific spatial dependences are considered in the plasma frequency, other accelerating solutions can be constructed, such as sinusoidal acceleration, or
 even obtaining arbitrary accelerations for each part of the wavepacket.

In order to demostrate the different accelerating behavior of an electromagnetic plasma wave, let us start describing the plasma as a  (special) relativistic two-fluid electron-positron plasma. Each plasma fluid (described by the subindex $i=e,p$) follow a dynamics described by the momentum equation
\begin{equation}
\partial_\nu{T_i^{\mu\nu}}= q_i n_i F^{\mu\nu}U_{\nu i}\, ,
\label{GRenmomen}
\end{equation}
where $U_i^{\mu}$ (such that $U_i^\mu U_{\mu i}=-1$) is the four velocity of each plasma fluid $i=p,e$,
with invariant charge $q_p=e=-q_e$ (with the electron charge $e$), and electron mass $m$ of each fluid element ($\partial^\mu$ is the four-derivative). The plasma is interacting with
an electromagnetic field $F_{\mu\nu}$.  The energy-momentum tensor  for an ideal plasma  is defined  as
$T^{\mu\nu}_i=m n_i f_i U^\mu_i U^\nu_i+p_i \eta^{\mu\nu}$ \cite{mah 03,Bek},
in terms of the the scalar number density $n_i$ (the rest frame density),  the scalar pressure $p_i$,  and enthalpy density $h_i=m n_i f_i$. Also, $\eta^{\mu\nu}=(-1,1,1,1)$ is the flat-spacetime metric.  Here, $f_i$ is a function of temperature $T_i$.
 In the special case of a relativistic
Maxwell distribution becomes $f= K_3(m /T)/ K_2(m /T)$ where $K_j$ is the modified Bessel functions of order $j$ (the Boltzmann constant is chosen  $k_B=1$).
Also, each plasma fluid must follow a continuity equation  
 \begin{equation}\label{conTnU}
\partial_\mu\left(n_iU^\mu_i\right)=0\, ,
\end{equation}
and couple to Maxwell equations
\begin{equation}
\partial_\nu {F^{\mu\nu}}=4\pi \sum_i q_i n_i U_i^\mu\, .
 \label{Maxwellcurved}
\end{equation}

The important issue about the previous equations is that they allow a unified dynamics, where the plasma fluid and electromagnetism are treated equally in terms of their vorticities
Thus, the  global dynamics can be described in terms of a unified  field tensor  \cite{mah 03, asenjo1,Bek} for each plasma fluid 
\begin{equation}\label{Mu}
M_i^{\mu\nu}=F^{\mu\nu}+\frac{m_i}{q_i}S_i^{\mu\nu}\ ,
\end{equation}
in which all kinematic and thermal (through $f$) aspects of the fluid are now contained by  the antisymmetric tensor $
S^{\mu\nu}=\partial^\mu\left(fU^{\nu}\right)-\partial^\nu\left(fU^{\mu}\right)$ for each fluid \cite{mah 03, asenjo1,Bek}.
In this way, Eq.~\eqref{GRenmomen} can be put in the form
\begin{equation}
q_i\ U_{\nu i} M_i^{\mu\nu}=-T_i\partial^\mu\sigma_i\, .
\label{eqmo}
\end{equation}
where $\sigma$ is the scalar entropy density of the fluid, and it is related to pressure through ${nT}\partial^\mu \sigma={mn \partial^\mu f-\partial^\mu p}$.
Furthermore, by the  antisymmetry of $M_{\mu\nu}$, each plasma fluid is  isentropic,  $U_{\mu i}\partial^\mu\sigma_i=0$.

Now, let us consider a cold plasma approximation, such as $\partial^\mu \sigma\approx 0$ (homentropic plasma), and $f\approx 1$. Then, Eq.~\eqref{eqmo} becomes $U_\nu M^{\mu\nu}=0$. In this case, the simplest solution is 
$M^{\mu\nu}=0$, that translate into $A^\mu=(m/q) U^\mu$ for each fluid. Using this solution of momentum equation into Maxwell
equations \eqref{Maxwellcurved}, we  obtain 
\begin{equation}
\partial_\nu {F^{\mu\nu}}=\omega_p^2 A^\mu\, ,
\label{geneMAxweF}
\end{equation}
 with $\omega_p^2=2\omega_{pe}^2$, where
$\omega_{pe}=\sqrt{4\pi e^2 n/m}$ is the electron plasma frequency. Here, we have assumed that the two fluids have the same density. 

Notice that, so far, no space or time functionality  dependence on density is required to obtain these transversal electromagnetic modes, satisying the Lorentz gauge $\partial_\mu A^\mu=0$.
Thus, Eq.~\eqref{geneMAxweF} can be written for the electromagnetic vector potential
to thus describe the relativistic dynamics of electromagnetic plasma waves. We chose, for sake of simplicity, a vector potential $A^\mu=(0,0,0,A(t,x,y))$, that satisties the Lorentz gauge. In this case, Eq.~\eqref{geneMAxweF} reduces to
\begin{equation}
    \left(\frac{\partial^2}{\partial t^2} - \frac{\partial^2}{\partial x^2} -\frac{\partial^2}{\partial y^2} +{\omega_p^2} \right){A}=0\, ,
    \label{densityequation}
\end{equation}

This the the main equation to explore in this work. It is enough to show that it predicts new different forms of accelerated propagation due to the two-dimensional space extension of the wave dynamics. Any other more complex electromagnetic plasma waves will contain similar effects.

In general, it is very well-known that 
 Eq.~\eqref{densityequation}
 can be solved in term of plane waves $A\sim \exp(i\omega t-i k_x x-i k_y y)$, with the dispersion relation $\omega^2=k_x^2+k_y^2+\omega_p^2$. This mode of propation has a constant velocity as it propagate in space and time.

Differently to the above, in this work we show that 
 Eq.~\eqref{densityequation} has other kind of propagating solutions that present acceleration depeding on the space-time dependence of the plasma frequency of the system. In the following sections, we explore three case in which the system can be solved exactly, to later in the conclusions to discuss the implications of these findings.

 \section{Schr\"odinger-like equation for two-dimensional electromagnetic plasma waves}

The accelerated properties of these plasma waves
can be explictly exposed  by putting Eq.~\eqref{densityequation} in its Schr\"odinger-like form. Let us introduce the light-cone coordinates  $\eta=x-t$, and  $\xi=x+t$ \cite{imb}. Then, let us assume that the  vector potential has the form
\begin{equation}
   A(t, x,y)=  \zeta(\eta, { y})\exp\left(ik \xi  \right)\, ,
    \label{exactnairprogrpab}
\end{equation}
where $k$ is an arbitrary constant.
Using this in Eq.~\eqref{densityequation} , we readily find the 
Schr\"odinger-like equation
\begin{equation}
   -4 ik\frac{\partial \zeta}{\partial \eta}+\frac{\partial^2\zeta}{\partial { y}^2}+\omega_p^2\zeta=0\, .
\label{schorec}
\end{equation}

It is well-know that this equation has accelerating solutions \cite{berry}. Any possible acceleration now occurs in the $\eta$--$y$ space. Also, notice that the plasma frequency acts like an effective potential to this equation. Depending on how it departs from a constant, different kind of accelerations can be obtained.

Thus, the information of the intensity of the electromagnetic plasma wave $A$ is now only contained in $\zeta$.

\section{Airy form of accelerated propagation}

Let us consider the simplest form of accelerated solution. This is obtained when plasma frequency has a linear dependence of the spatial dimension $y$ (which is transvesal to the light-cone) such that 
\begin{equation}
\omega_p^2(y)=\alpha+\beta y\, ,
\end{equation}
where $\alpha$ and $\beta$ are constants. In this case, we find the accelerating solution in terms by Airy functions ${\mbox{Ai}}$ of Eq.~\eqref{schorec}, as \cite{berry}
\begin{eqnarray}
\zeta(\eta,y)&=&{\mbox{Ai}}\left[\left(8 a k^2-\beta\right)^{1/3}\left(y - \frac{a}{2}\eta^2\right)\right]\times\nonumber\\
&&\exp\left[-2i a k \eta\left( y- \frac{(16 a k^2-\beta)}{48 k^2}\eta^2 \right) -i\frac{\alpha}{4k}\eta \right] \, ,\nonumber\\
&&
\label{airysolution}
\end{eqnarray}
where $a$ is arbitrary, representing the acceleration of these wavepacket. Each part of the wavepacket follows a curved trayectory given by $y - {a}\eta^2/2={\mbox{constant}}$, as it can be seen through the amplitude of the wave $|A|={\mbox{Ai}}\left[\left(8 a k^2-\beta\right)^{1/3}\left(y - {a}\eta^2/2\right)\right]$.
 In this case, the acceleration is constant, as it can be obtained from $d^2 y/dt^2=d^2 y/d\eta^2=a$. 

\begin{figure}
  \centering
\includegraphics[width = 3.1in]{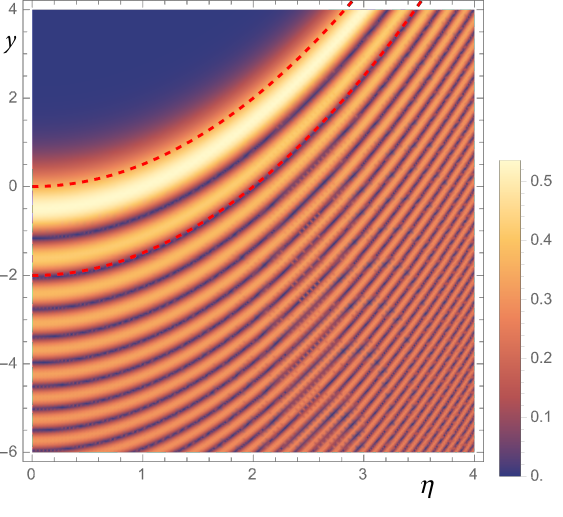}
\caption{Density plot for magnitude of solution \eqref{airysolution}, with $8 ak^2-\beta=8$, and $a=1$, to show the intensity of the accelerated wave. 
 Red dashed lines corresponds to parabolic trajectories $y-\eta^2/2=\xi_0$, with $\xi_0= -2,\, 0$.}
\label{figuraAiry}
\end{figure}

For instance, this can be seen in Fig.~\ref{figuraAiry}, where the magnitude of solution \eqref{airysolution} is  plotted for $8 ak^2-\beta=8$, and $a=1$. 
Each part of the wavepacket experiences acceleration,
 following curved trayectories in the $\eta$--$y$ space. To explicitly show this, we show the trayectories $y-\eta^2/2=0$, and $y-\eta^2/2=-2$ in red dashed lines. Every part of the wavepacket follow the same trayectories.

\section{Weber form of accelerated propagation}
 
In the case that plasma frequency have a quadratic dependence on the spatial direction $y$ (where $\alpha$ and $\beta$ are again constants)
\begin{equation}
\omega_p^2(y)=\alpha-\beta y^2\, ,
\end{equation}
then other form of accelerated propagation is present. In those case, solution of Eq.~\eqref{schorec} is
\begin{eqnarray}
\zeta(\eta,y)&=&{\mbox{W}}\left[\chi(\eta,y)\right]\exp\left[i \beta^{1/4} y \cos\left( \frac{\sqrt{\beta}}{2k}\eta\right)
-i\mu(\eta)\right] \, ,\nonumber\\
&&
\label{webersoltuoin}
\end{eqnarray}
where 
\begin{eqnarray}
\chi(\eta,y)&=&\beta^{1/4} y +\sin\left( \frac{\sqrt{\beta}}{2k}\eta\right)\, ,\nonumber\\
\mu(\eta)&=& \frac{\alpha}{4k}\eta-\frac{\sqrt{\beta}\, n}{4 k}\eta- \frac{1}{4}\sin\left( \frac{\sqrt{\beta}}{2k}\eta\right)\, ,
\end{eqnarray}
with $n$ as a integer number.
Here $W$ is the solution of the Weber equation $W''(\chi)+(n-\chi^2)W(\chi)=0$, which can be written in terms of parabolic cylinder function $D_{(n-1)/2}(\sqrt{2}\chi)$. 

Notice that the acceleration in this case is not constant. Each part of the wavepacket accelerated in trayectories given by $\chi={\mbox{constant}}$. Thus, the acceleration of those parts is given by $d^2y/dt^2=d^2 y/d\eta^2=\beta^{3/4}\sin\left( {\sqrt{\beta}}\eta/{2k}\right)/(4 k^2)$. So, the wavepacket oscillates while it accelerates.

\begin{figure}
  \centering
\includegraphics[width = 3.1in]{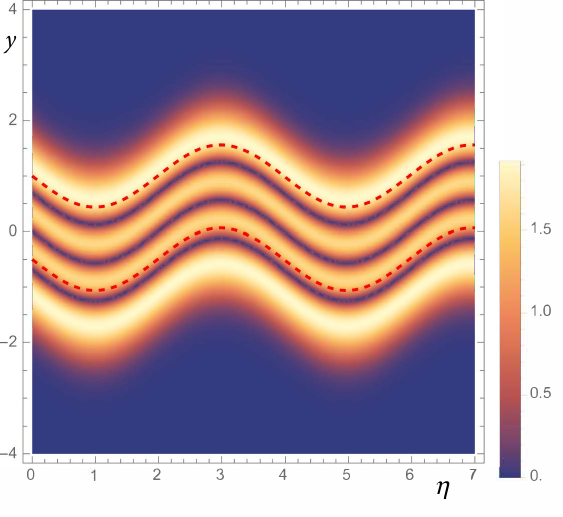}
\caption{Density plot for magnitude of solution \eqref{webersoltuoin}, with $n=7$, $\beta=10$, and $k=1$. The non-constant acceleration is shown explicitly.
 Red dashed lines corresponds to parabolic trajectories $y-10^{-1/4} \sin(\sqrt{10} \eta/2)=\xi_0$, with $\xi_0= -1/2,\, 1$.}
\label{figuraWeber}
\end{figure}

In Fig.~\ref{figuraWeber}, we display a density plot for magnitude of solution \eqref{webersoltuoin} for $n=7$. The solution is displayed for $\beta=10$, and $k=1$. The sinusoidal acceleration of each part of the wavepaket is explicitly shown in the $\eta$--$y$ space. As an example, in red dashed lines, we also show the parabolic trayectories
  $y-10^{-1/4} \sin(\sqrt{10} \eta/2)= -1/2$, and $y-10^{-1/4} \sin(\sqrt{10} \eta/2)= 1$.

\section{Arbitrary form of accelerated propagation}

More general and arbitrary forms of acceleration for the electromagnetic plasma wavepackets can be obtained in a simpler way if the plasma frequency has a $\eta$-dependence as
\begin{equation}
\omega_p^2(\eta,y)=F(\eta)\left(\alpha+\beta y\right)\, ,
\end{equation}
where anew $\alpha$ and $\beta$ are constants, and $F$ is an arbitrary function only of $\eta$. This implies that the plasma frequency has a modulation along a light-cone coordinate. This can occur, for example,  in a luminal plasma-vacuum interface (luminal plasma slab) \cite{bulanov,bulanov2}.

In this case, the
 solution of Eq.~\eqref{schorec} is again in term of   Airy functions, given by
\begin{eqnarray}
\zeta(\eta,y)&=&{\mbox{Ai}}\left[\chi(\eta,y)\right]\exp\left[i \Theta(\eta) y+i\Gamma(\eta) \right] \, ,
\label{airysolutionGeneral}
\end{eqnarray}
where
\begin{eqnarray}
\chi(\eta,y)&=&\chi_0\,  y+\gamma(\eta)\, ,\nonumber\\
\gamma(\eta)&=&\frac{\chi_0}{2k}\int \Theta d\eta\, ,\nonumber\\
\Theta(\eta)&=&\frac{\chi_0^3}{4 k}\eta-\frac{\beta}{4k}\int F d\eta\, ,\nonumber\\
\Gamma(\eta)&=& \frac{\chi_0^2}{4 k}\int\gamma d\eta+\frac{1}{4k^2}\int\Theta^2 d\eta-\frac{\alpha}{4k}\int F d\eta\, \nonumber\\
&&
\end{eqnarray}
where $\chi_0$ is an arbitrary constant. 

For this  electromagnetic plasma mode, each part of the wavepacket accelerates as $d^2y/dt^2=d^2 y/d\eta^2=-(1/2k)\, d\Theta/d\eta=(\beta F-\chi_0^3)/(8 k^2)$, which is completly dependent of the form of $F$, and therefore, arbitrary.

\section{Conclusion}

Relativistic electromagnetic plasma waves can be studied for   plasma frequencies that can have, in general, spatial and temporal dependence.
Therefore, various forms of different accelerating solutions are possible for the linear equation \eqref{densityequation}. 

Besides of the simple Airy accelerating mode \eqref{airysolution} (for constant or linear spatial dependence of the plasma frequency), or the Weber accelerating mode \eqref{webersoltuoin} (for a quadratic spatial dependence), other possible modes can be found for more complex spatial dependence of the plasma frequency. For instante, for a cubic spatial dependence, another, more complicated forms of the accelerating solution can be constructed. 
Similarly, a Weber-like solution with arbitrary acceleration, can be also found in the case of quadratic spatial dependence of the plasma frequency [a generalization of solution \eqref{airysolutionGeneral}]. 

Therefore, as a conclusion, the purpose of this work is to show that all these (and more) different accelerated solutions are consequence of the constant and  non-constant plasma frequency. Electromagnetic plasma wavepackets (and any other wave equation with similar form) predicts the above accelerated solution, which  can also be generalized in order to produce normalizable solutions
\cite{lekner}. 

These above accelerating  relativistic electromagnetic plasma wave
solutions open unexplored forms of propagation. They can have, for instance, very interesting impact in new forms of laser-plasma interaction \cite{Efremidis,JianXingLi, takale} where different forms of accelerating laser beams, with diverse trayectories, can be engineered by appropriately modulating the plasma frequency of the plasma medium. This is left for future studies.


\begin{thebibliography}{}



\bibitem{neomi} N. Wiersma, N. Marsal, M. Sciamanna and D. Wolfersberger, Sci. Rep. {\bf 6},35078 (2016).

\bibitem{Jiang} Y. Jiang, K. Huang, and X. Lu,  Opt. Express {\bf 20}, 18579 (2012).

\bibitem{abdo} D. Abdollahpour {\it et al.}, Phys. Rev. Lett. {\bf 105}, 253901 (2010).

\bibitem{bouch} T. Bouchet {\it et al.} Sci Rep {\bf 12}, 9064 (2022). 

\bibitem{chong} A. Chong {\it et al.}, Nature Photon {\bf 4}, 103 (2010). 

\bibitem{Kaminer} I. Kaminer {\it et al.}, Phys. Rev. Lett. {\bf 108}, 163901 (2012).

\bibitem{panag} P. Panagiotopoulos {\it et al.},  Nat Commun {\bf 4}, 2622 (2013).

\bibitem{esat} H. Esat Kondakci and A. F. Abouraddy, Phys. Rev. Lett. {\bf 120}, 163901 (2018).


\bibitem{Nikolaos} N. K. Efremidis {\it et al.}, Optica {\bf 6}, 686 (2019).

\bibitem{moya} S. Ch\'avez-Cerda {\it et al.}, Opt. Exp. {\bf 19}, 16448 (2011).
\bibitem{Baumgartl} J. Baumgartl, M. Mazilu and K. Dholakia, Nature Photon {\bf 2}, 675 (2008).


\bibitem{water} S. Fu, Y. Tsur, J. Zhou, L. Shemer and A. Arie,
Phys. Rev. Lett. {\bf 115}, 034501 (2015).

\bibitem{zhao} S. Zhao {\it et al.}, Sci Rep {\bf 4}, 6628 (2014). 

\bibitem{chencho} D.-C. Chen {\it et al.}, Appl. Phys. Lett. {\bf 114}, 053504 (2019).



\bibitem{bookolivier} O. Vall\'ee and M. Soares, {\it Airy functions and applications to physics} (Imperial College Press, 2004).

\bibitem{asenjoheat} F. A. Asenjo and S. A. Hojman, Eur. Phys. J. Plus {\bf 136}, 677 (2021).

\bibitem{asenjoheat2}  F. A. Asenjo and S. A. Hojman, Eur. Phys. J. Plus {\bf 137}, 1201 (2022).

\bibitem{asenjograv} F. A. Asenjo and S. A. Hojman, Eur. Phys. J. C {\bf 81}, 98 (2021).


\bibitem{asenjoelectron} M. A. Winkler, C. V\'asquez-Wilson and F. A. Asenjo, Eur. Phys. J. D  {\bf 77}, 97 (2023).

\bibitem{YChen}  Y. Chen, L. Wu, Y. Liu, Y. Wu, Z. Lin and D. Deng, Phys. Plasmas {\bf 27}, 082104 (2020).

\bibitem{LWu} L. Wu {\it et al.}, Phys. Plasmas {\bf 26}, 092111 (2019).


\bibitem{li} H. Li, X. Li and J. Wang, J. Plasma Phys. {\bf 82}, 905820103 (2016).

\bibitem{klimo} A. E. Minovich {\it et al.}, Laser Photon. Rev. {\bf 8}, 221 (2014).

\bibitem{asenjoelectron2} F. A. Asenjo, J. Plasma Phys. {\bf 90}, 905900119 (2024).





\bibitem{mah 03} S. M. Mahajan, Phys. Rev. Lett. {\bf 90}, 035001 (2003).
\bibitem{Bek}  J. D. Bekenstein, Astrophys. J. {\bf 319}, 270 (1987).

\bibitem{asenjo1} F. A. Asenjo, S. M. Mahajan and A. Qadir, Phys. Plasmas {\bf 20}, 022901 (2013).





\bibitem{imb} I. M. Besieris, A. M. Shaarawi and R. W. Ziolkowski, Am. J. Phys. {bf 62},
519 (1994).





\bibitem{berry} M. V. Berry and N. L. Balazs,  Am. J. Phys. {\bf 47}, 264 (1979).


\bibitem{bulanov} A. Zhidkov, T. Esirkepov, T. Fujii, K. Nemoto, J. Koga and S. V. Bulanov,
Phys. Rev. Lett. {\bf 103}, 215003  (2009).

\bibitem{bulanov2}   T. Z. Esirkepov and S. V. Bulanov,
Phys. Rev. E {\bf 109}, L023202  (2024).


\bibitem{lekner} J. Lekner, Eur. J. Phys. {\bf 30}, L43 (2009). 

\bibitem{Efremidis} N. K. Efremidis, Z. Chen, M. Segev and D. N. Christodoulides, Optica {\bf 6}, 686  (2019).

\bibitem{JianXingLi} J.-X. Li, W.-P. Zang and J.-G. Tian, Opt. Express {\bf 18}, 7300 (2010).

\bibitem{takale} V. S. Pawar, S. R. Kokare, S. D. Patil and M. V. Takale,  Laser and Particle Beams {\bf 38}, 204 (2020).





\end{thebibliography}
\end{document}